\documentclass[pre,twocolumn,citesort,aps,letterpaper]{revtex4}%
\usepackage{amsfonts}
\usepackage{amsmath}
\usepackage{amssymb}
\usepackage{graphicx}%
\providecommand{\U}[1]{\protect\rule{.1in}{.1in}}

\begin{document}
\title[Steric effects in the dynamics of electroyltes I]{Steric effects in the dynamics of electrolytes at large applied voltages:\\I. Double-layer charging}
\author{Mustafa Sabri Kilic}
\author{Martin Z. Bazant}
\affiliation{Department of Mathematics, Massachusetts Institute of Technology, Cambridge,
MA 02139,USA.}
\author{Armand Ajdari}
\affiliation{Labortoire de Physico-Chimie Theorique, UMR ESPCI-CNRS 7083, 10 rue Vauquelin,
F-75005 Paris, France.}
\pacs{PACS number}

\begin{abstract}
The classical Poisson-Boltzmann (PB) theory of electrolytes assumes a dilute
solution of point charges with mean-field electrostatic forces. Even for very
dilute solutions, however, it predicts absurdly large ion concentrations
(exceeding close packing) for surface potentials of only a few tenths of a
volt, which are often exceeded, e.g. in microfluidic pumps and electrochemical
sensors. Since the 1950s, several modifications of the PB equation have been
proposed to account for the finite size of ions in equilibrium, but in this
two-part series, we consider steric effects on diffuse charge dynamics (in the
absence of electro-osmotic flow). In this first part, we review the literature
and analyze two simple models for the charging of a thin double layer, which
must form a condensed layer of close-packed ions near the surface at high
voltage. A surprising prediction is that the differential capacitance
typically varies non-monotonically with the applied voltage, and thus so does
the response time of an electrolytic system. In PB theory, the capacitance
blows up exponentially with voltage, but steric effects actually cause it to
decrease above a threshold voltage where ions become crowded near the surface.
Other nonlinear effects in PB theory are also strongly suppressed by steric
effects: The net salt adsorption by the double layers in response to the
applied voltage is greatly reduced, and so is the tangential "surface
conduction" in the diffuse layer, to the point that it can often be neglected
compared to bulk conduction (small Dukhin number).

\end{abstract}
\volumeyear{2006}
\volumenumber{number}
\issuenumber{number}
\eid{identifier}
\date[Date text]{date}
\received[Received text]{date}

\revised[Revised text]{date}

\accepted[Accepted text]{date}

\published[Published text]{date}

\maketitle



\section{ Introduction}

In this two-part series, we develop a simple analytical theory for the
dynamics of electrolytes, taking into account steric effects of finite ion
size. Motivated by recent experiments in microfluidics, microbatteries, and
electrochemical sensors, our motivation is to describe the response of an
electrolyte to an applied voltage of several volts, which is large enough to
cause crowding of ions near a surface, even if the bulk solution is very
dilute and in the absence of surface reactions. The ions thus lose their
classical Poisson-Boltzmann distribution, which has major implications for
their dynamics.

As a guide to the reader, we summarize the main results. The present "Part I"
begins in this section with a historical review of dilute solution theory, its
limitations at large voltages, and attempts to account for steric hindrance,
specific interactions, and many-body electrostatics. As a first approximation,
we focus only on steric effects, and analyze the dynamical response of a thin
diffuse layer to a large applied voltage, using two simple continuum models
(section~\ref{sec:models}). The key results, common to both steric models,
are: (i) the diffuse layer's differential capacitance is bounded and decreases
at large voltage (section~\ref{sec:circuit}), and (ii) it cannot easily engulf
enough ions to perturb the bulk concentration or to conduct significant
tangential currents (section~\ref{sec:beyond}). These predictions are
completely opposite to those of dilute solution theory (based on the
Gouy-Chapman model of the diffuse layer). In the companion paper "Part II", we
propose general, time-dependent equations with steric constraints and revisit
the parallel-plate charging problem in more detail.

\subsection{ Dilute solution theory}

For the past century, dilute solution theory has provided the standard model
of electro-diffusion of
ions~\cite{rubinstein_book,newman_book,lyklema_book_vol1} and electrokinetic
phenomena~\cite{lyklema_book_vol2,hunter_book}. The fundamental assumption is
that the chemical potential of a point-like ion $i$ in a dilute solution has
the simple form,
\begin{equation}
\mu_{i} = kT \ln c_{i} + z_{i}e \psi\label{eq:mu_dilute}%
\end{equation}
where $z_{i} e$ is the charge, $c_{i}$ the concentration, and $\psi$ the
electrostatic potential, determined in a mean-field approximation by Poisson's
equation,
\begin{equation}
-\nabla\cdot(\varepsilon\nabla\psi) = \rho= \sum_{i} z_{i} e c_{i}.
\label{eq:poisson}%
\end{equation}
typically with a constant permittivity $\varepsilon$. The form
(\ref{eq:mu_dilute}) is sometimes called the ``ideal'' component of the
chemical potential~\cite{gillespie2002}, to which various ``excess''
components at finite concentration can be added (see below).

In many situations, it is assumed that the ions are in quasi-thermal
equilibrium with a Boltzmann distribution,
\begin{equation}
c_{i}=c_{i}^{0}e^{-z_{i}e\psi/kT} \label{eq:bolt}%
\end{equation}
with a reference concentration $c_{i}^{0}$, in which case
Equation~(\ref{eq:poisson}) reduces to the Poisson-Boltzmann equation (PB).
For example, to analyze a thin double layer, it is natural to choose
$c_{i}^{0}$ to be the concentration of species $i$ in the nearby neutral bulk
solution. In most situations, the PB equation is hard enough to solve that the
the Debye-H\"{u}ckel linearization for small potentials, $|\psi|\ll\psi_{T}$,
is required for analytical progress, where $\psi_{T}=kT/z_{i}e$ is the thermal
voltage ($\approx25$ mV for monovalent ions at room temperature).

The well known exception, which admits a simple solution, is the case of a
symmetric binary $z:z$ electrolyte in a 1D geometry, where the PB equation in
the form
\begin{equation}
\varepsilon\frac{\partial^{2} \psi}{\partial x^{2}} = 2zec_{0} \, \sinh\left(
\frac{ze\psi}{kT}\right)  \label{eq:pbsym}%
\end{equation}
was solved analytically by Gouy~\cite{gouy1910} and Chapman~\cite{chapman1913}
(GC) for a semi-infinite electrolyte of bulk concentration $c_{0}=c_{+}%
^{0}=c_{-}^{0}$ near a flat charged
surface~\cite{newman_book,lyklema_book_vol1,hunter_book}. (It is less well
known that Gouy~\cite{gouy1910} also solved the case of an asymmetric $2z:z$
electrolyte.) Although this solution may seem quite special, it describes the
most common situation where diffuse charge is confined to a thin
capacitor-like ``double layer'' near a solid surface. The width of the diffuse
layer is the Debye~\cite{debye1928} (or, more properly, Gouy~\cite{gouy1910})
screening length,
\begin{equation}
\lambda_{D}=\sqrt{\frac{\varepsilon kT}{2z^{2}e^{2}c_{0}}} \label{debye}%
\end{equation}
as can be seen by scaling (\ref{eq:pbsym}) with $\tilde{x}=x/\lambda_{D}$ and
$\tilde{\psi}=\psi/\psi_{T}$ to obtain the dimensionless form, $\tilde{\psi
}^{\prime\prime} = \sinh\tilde{\psi}$. The screening length ($\lambda
_{D}\approx1-100$ nm in aqueous electrolytes) is typically much smaller than
any geometrical length, so bulk solution remains quasi-neutral with diffuse
charge confined to thin, locally flat, quasi-equilibrium double layers.

Due to its analytical simplicity and wide applicability, the Gouy-Chapman
solution has become the standard model for highly charged double layers. It is
the basis for the classical theory of tangential ``surface conduction''
through the diffuse
layer~\cite{bikerman1933,bikerman1940,deryagin1969,dukhin1993}, as well as the
recently noted phenomenon of salt ``adsorption'' (or ``uptake'') from the
neutral bulk by the diffuse layer in response to a large applied
voltage~\cite{bazant2004,chu2006}. Such predictions of the GC model have major
implications for electrokinetic phenomena of the first kind, such as
electro-osmosis, electrophoresis, streaming potential, and diffusiophoresis,
at large surface potentials~\cite{lyklema_book_vol2,hunter_book}.

Dilute solution theory has also been used in nearly every model of diffuse
charge during the passage of current. Near equilibrium, the flux density of
ion $i$ is proportional to the gradient of its chemical potential
(\ref{eq:mu_dilute}),
\begin{equation}
F_{i} = -b_{i} c_{i} \nabla\mu_{i} = -D_{i} \left(  \nabla c_{i} + \frac
{z_{i}e}{kT} c_{i} \nabla\psi\right)
\end{equation}
where Einstein's relation, $D_{i} = b_{i}/kT$, relates the ion's mobility
$b_{i}$ to its diffusivity $D_{i}$. For a system in quasi-steady state,
$\nabla\cdot F_{i} = 0$, the nonzero current, $J = \sum_{i} z_{i} e F_{i}$,
only slightly perturbs the Boltzmann distribution of the ions. As a result,
the GC solution is also widely used to describe diffuse-layer effects on
electrode reaction kinetics~\cite{newman_book,bazant2005} (the Frumkin
correction~\cite{frumkin1933}), up to Nernst's limiting current, where the
bulk concentration of an electro-active species vanishes. At larger
``superlimiting'' currents, the PB equation loses validity, but
dilute-solution theory is still used to describe diffuse charge, which loses
its Boltzmann distribution~\cite{smyrl1967} and extends into the bulk as
``space charge''~\cite{rubinstein1979}, while retaining an inner boundary
layer at the screening length~\cite{chu2005}. The dilute-solution space-charge
model is the basis for theories of electrokinetic phenomena of the second
kind, such as super-fast electrophoresis~\cite{dukhin1991} and hydrodynamic
instability at a limiting current~\cite{rubinstein2000}.

Dilute solution theory has also been used to describe the dynamics of
electrolytes, subject to time-dependent applied voltages. The classical
description comes from the Poisson-Nernst-Planck equations (PNP), which
consist of Eq.~(\ref{eq:poisson}) and mass conservation laws for the ions,
\begin{equation}
\frac{\partial c_{i}}{\partial t} = - \nabla\cdot F_{i} = D_{i} \left[
\nabla^{2} c_{i} + \frac{z_{i}e}{kT} \nabla\cdot(c_{i} \nabla\psi)\right]
\end{equation}
where $D_{i}=$ constant is normally assumed. Again, general solutions are only
possible upon linearization, where the potential satsifies the
Debye-Falkenhagen equation~\cite{debye1928}. In the nonlinear regime, for thin
double layers, the GC solution again holds since the diffuse-charge remains in
quasi-equilibrium, and the diffuse-layer acts like a voltage-dependent
differential capacitance in series with a bulk resistor~\cite{macdonald1954_b}%
. Such equivalent circuit models can be derived systematically from the PNP
equations by aymptotic analysis, which also reveals corrections at large
voltages, due to bulk diffusion in response to salt
adsorption~\cite{bazant2004} and surface conduction~\cite{chu2006} at large
applied voltages. Such nonlinear diffuse-layer effects could be important in
intepretting impedance spectra in electrochemical
sensing~\cite{macdonald1990,geddes1997}, or in understanding high-rate
thin-film rechargeable
batteries~\cite{dudney1995,wang1996,neudecker2000,takami2002}.

Another current motivation to study nonlinear diffuse-charge dynamics comes
from ``induced-charge'' electrokinetic phenomena~\cite{iceo2004a}. The
preceding models of double-layer relaxation have been used extensively in
theories of AC pumping of liquids over electrode
arrays~\cite{ramos1999,ajdari2000,gonzalez2000,brown2001,ramos2003,bazant2006}%
, induced-charge electro-osmotic flows around metallic
colloids~\cite{gamayunov1986,murtsovkin1996} and
microstructures~\cite{iceo2004b,levitan2005,bazant2006}, and
dielectrophoresis~\cite{shilov1981,simonova2001,squires2006} and
induced-charge
electrophoresis\cite{iceo2004a,yariv2005,squires2006,saintillan2006} of
polarizable particles in electrolytes, although a nonlinear analysis based on
the PNP equations has not yet been attempted for any of these situations. This
may be a good thing, however, since we will show that dilute solution theory
generally breaks down in the regime of experimental interest, where the
applied voltage is much larger than the thermal voltage, $V > \psi_{T}$.

\subsection{Validity of the nonlinear model for dilute solutions}

Dilute solution theory provides a natural starting point to understand
nonlinear effects in electrolytes, and the GC model is used in all of the
examples above to model the diffuse layer at large applied potentials. In
spite of its mathematical appeal, however, nonlinear dilute solution theory
has limited applicability, due to the exponential sensitivity of the
counter-ion concentration to voltage. For example, in the typical case of
quasi-equilibrium, the Boltzmann factor (\ref{eq:bolt}) brings nonlinearity to
the PB equation (\ref{eq:pbsym}) when $\psi> \psi_{T}$, but this dependence is
so strong that the surface potential cannot get much larger without
invalidating the assumption of a dilute solution.

To see this, note that there must be a maximum concentration, $c_{max}=a^{-3}%
$, of counterions packed with typical spacing $a$ near a highly charged
surface. This effective ion size is clearly no smaller than the ionic radius
(typically $\approx1\mathring{A}$), although it could be considerably larger
(several nm), taking into account hydration effects and ion-ion correlations,
especially in a large electric field. For the sake of argument, consider a
very dilute bulk solution, $c_{0}=10^{-5}$M, with $a=3\mathring{A}$ and $z=1$.
The nonlinear regime begins at a diffuse-layer potential $\Psi_{D}$ $\approx
kT/ze=25$ mV, but the steric limit is reached in dilute solution theory at
$\Psi_{c}\approx13kT/ze=330$ mV, where
\begin{equation}
\Psi_{c}=-\frac{kT}{ze}\ln(a^{3}c_{0})=\frac{kT}{ze}\ln\left(  \frac{c_{max}%
}{c_{0}}\right)  .
\end{equation}
Since the solution ceases to be \textquotedblleft dilute\textquotedblright%
\ well below the steric limit, there is only a narrow window of surface
potentials ($\approx25-200$ mV), sometimes called \textquotedblleft weakly
nonlinear regime\textquotedblright~\cite{bazant2004}, where nonlinearity
arises and the fundamental assumption (\ref{eq:mu_dilute}) remains valid.

Unfortunately, the most interesting predictions of the non-linear PB theory
tend to be in the \textquotedblleft strongly nonlinear
regime\textquotedblright, where the dilute approximation fails dramatically.
For example, the Dukhin number, which controls the relative importance of
tangential conductivity in a thin diffuse layer, $\sigma_{s}$, compared to
bulk conductivity, $\sigma$, at a geometrical scale $L$, has the following
form~\cite{bikerman1933,bikerman1940,deryagin1969},
\begin{equation}
\mbox{Du}=\frac{\sigma_{s}}{\sigma L}=4\frac{\lambda_{D}}{L}\sinh^{2}\left(
\frac{ze\Psi_{D}}{4kT}\right)  , \label{eq:Du}%
\end{equation}
assuming the Gouy-Chapman model. This is also the dimensionless ratio,
$\Gamma_{s}/c_{0}L$, of the excess surface concentration of ions, $\Gamma_{s}%
$, relative to the bulk concentration, so it also governs the (positive)
adsorption of neutral salt from the bulk in response to an applied
voltage~\cite{bazant2004,lyklema2005}. The general derivation of
Eq.~(\ref{eq:Du}) assumes a thin double layer~\cite{chu2006}, but in that case
($\lambda_{D}\ll L$) a large Dukhin number corresponds to situations where the
steric constraint is significantly violated, $\Psi_{D}>\Psi_{c}$, rendering
Equation (9) inappropriate. Similar concerns apply to other nonlinear effects
in Gouy-Chapman theory, such as the rapid increase of the differential
capacitance (defined below) with surface potential,
\begin{equation}
C_{D}=\frac{\varepsilon}{\lambda_{D}}\cosh(\frac{ze\Psi_{D}}{2kT}).
\label{cdpb_2}%
\end{equation}
which would have important implications for electrochemical relaxation around
conductors~\cite{simonov1977,bazant2004,chu2006} and for AC
electro-osmosis~\cite{olesen2006, labchip2006}.

\subsection{ Beyond the Poisson-Boltzmann picture}

We are certainly not the first to recognize the limitations of dilute solution
theory. Historically, concerns about the unbounded capacitance of a thin
diffuse layer in the GC solution (\ref{cdpb_2}) first motivated
Stern~\cite{stern1924} to hypothesize that there must also be compact layer of
adsorbed ions on the surface, as originally envisioned by
Helmholtz~\cite{helmholtz1879}. The Stern layer is where electrochemical
reactions, such as ion dissociation (setting the equilibrium charge on a
dielectric) and/or redox couples (setting the Faradaic current at an
electrode), are believed to occur, within a molecular distance of the solid
surface~\cite{lyklema_book_vol1,lyklema_book_vol2}. The Stern layer
capacitance (see below) helps to relieve the overcharging of the diffuse layer
in Gouy-Chapman theory, but, due to steric constraints, it too cannot
conceivably withstand a voltage much larger than $\Psi_{c}$. Instead, at
larger voltages, the region of ion accumulation must inevitably extend away
from the surface into the solution, where ions undergo hindered transport in a
concentrated solution without having specific interactions with the solid.

The most basic aspect of a concentrated solution is the finite molecular
length scale, $a_{0}$. Over half a century ago, Wicke and
Eigen~\cite{wicke1952,freise1952,eigen1954} made perhaps the first attempts to
extend dilute solution theory to account excluded volume effects in a simple
statistical mechanical treatment. The theory was developed further in the past
decade by Iglic and
Kral-Iglic~\cite{iglic1994,kralj-iglic1996,bohinc2001,bohinc2002} and
Borukhov, Andelman, and Orland~\cite{borukhov1997,borukhov2000,borukhov2004}.
These authors proposed free energy functionals, based on continuum
(mean-field) approximations of the entropy of equal-sized ions and solvent
molecules, which they minimized to derive modified Poisson-Boltzmann equations
(MPB). The motivation for this work was mainly to address the effect of large
ions, whose sizes may be comparable to the screening length, in an equilibrium
diffuse layer with $\Psi_{D}\approx\psi_{T}$. Our main point here is that
crowding effects can also be very important for small ions near a polarizable
surface, when subjected to a \textquotedblleft large\textquotedblright%
\ voltage the exceeding the threshold $\Psi_{c}$. No matter how dilute is the
bulk solution, a sufficiently large electric field can always draw enough ions
to the surface to create a concentrated solution in the diffuse layer.

There have also been attempts to go beyond the mean-field approximation of
steric effects, by treating specific (and in some cases also Coulombic)
ion-ion and ion-wall interactions. Simple MPB equations have been proposed,
which modify the charge density to account for the spatial correlation
function between an ion in solution and a flat wall (via an effective external
potential) based on molecular dynamics
simulations~\cite{freund2002,qiao2003,joly2004}, but such models are not
easily extended to any other geometry, such as a rough wall~\cite{kim2006}.
Corrections of this type can also be obtained from a general probabilistic
model of interacting ions, whose dynamics are given by nonlinearly coupled
Langevin equations. Using this approach, Schuss, Nadler and Eisenberg
rigorously derived ``conditional PNP'' (and PB) equations where each ion
concentration $c_{i}(\mathbf{r})$ in the mean-field Poisson's equation
(\ref{eq:poisson}) is replaced by the conditional probability density of
finding an ion at a certain position, given the positions of the other
ions~\cite{schuss2001,nadler2003,nadler2004}, although a simple closure of the
model requires further assumptions about statistical correlations.

There are a variety of general statistical mechanical approaches from liquid
state theory~\cite{henderson_book}, which have been applied to electrolytes,
taking into account not only steric effects, but also many-body electrostatic
correlations (see below). Since the 1970s, the modest, but challenging, goal
has been to accurately predict the equilibrium distribution of ions in Monte
Carlo simulations of the \textquotedblleft primitive model\textquotedblright%
\ of charged hard spheres in a homogeneous dielectric continuum bounded by a
hard, charged wall. Typically, the model is studied in the limit of
\textquotedblleft small\textquotedblright\ surface potentials ($\Psi
_{D}\approx kT/e$) relevant for equilibrium surfaces. For example, a plethora
of MPB equations (such as \textquotedblleft MPB4\textquotedblright,
\textquotedblleft MPB5\textquotedblright,...) perturbing Gouy-Chapman theory
have been derived by variations on the Mean Spherical Approximation
~\cite{levine1978,outhwaite1980,carnie1981,outhwaite1982,outhwaite1983,carnie1984,bhuiyan2005}%
. More complicated, but often more accurate, theories have also been derived
using statistical Density Functional Theory
(DFT)~\cite{gillespie2003,gillespie2005,reszko2005}. By writing the ion flux
as the gradient of an electrochemical potential obtained from the DFT free
energy functional~\cite{gillespie2002} (as we do in Part II, using the much
simpler functional of Borukhov et al.~\cite{borukhov1997}), it has been shown
that the selectivity of ion channels can be understood to a large extent in
terms of entropic effects of charged hard spheres~\cite{roth2005}.

In spite of many successes and the appealing first-principles nature of
liquid-state theories, however, they are not ideally suited for our purpose
here of capturing time-dependent, large-voltage effects in a simple model.
Liquid-state theories are not analytically tractable and thus not easily
coupled to macroscopic continuum models, although perhaps empirical fits could
be developed for quantities of interest. For example, the DFT approach
requires nontrivial numerical methods, just to evaluate the steady flux in a
one-dimensional ion channel~\cite{gillespie2002,roth2005}. Another concern is
that liquid-state theories have been designed and tested only for
\textquotedblleft small\textquotedblright\ voltages ($\Psi_{D}\approx kT/e$)
at equilibrium surfaces, so it is not clear how well they would perform at
large, time-dependent voltages, since there are no rigorous error bounds. For
example, the double-layer differential capacitance in modified PB theories
often increases with voltage~\cite{outhwaite1980}, sometimes even faster than
in PB theory~\cite{carnie1981}, but we will argue below that it must decrease
at large voltages, especially in concentrated solutions. In light of the
scaling $e\Psi_{D}/kT$ from PB theory (\ref{eq:bolt}), a related problem is
that, until recently~\cite{reszko2005,bhuiyan2005}, most theories have been
unable to predict the decay of capacitance at low temperature in concentrated
solutions. Although liquid-state theories for large voltages should certainly
be developed, we will focus on much simpler mean-field, continuum descriptions
of steric effects, in the hope of at least capturing some qualitative features
of nonlinear dynamics analytically, across a large range of applied voltages

Consistent with this approach, we will also work with the mean-field continuum
description of electrostatic interactions (\ref{eq:poisson}), which neglects
discrete many-body correlations. In passing, we point the reader to some of
the extensive literature on corrections to PB electrostatics, for the simplest
possible model of an equilibrium liquid of point-like charges in a uniform
dielectric medium. In the absence of specific interactions (such as steric
repulsion), the fundamental length scale for ion-ion correlations is the
Bjerrum length ($\approx7\mathring{A}$ in water at room temperature),%
\[
l_{B}=\frac{e^{2}}{4\pi\varepsilon kT}%
\]
at which the bare Coulomb energy balances the thermal energy. Interesting
consequences of many-body electrostatics, not present in the mean-field PB
description, include Oosawa-Manning condensation of counterions around charged
rods in polyelectrolytes~\cite{oosawa_book,oosawa1968,manning1969},
renormalized charge of colloidal
spheres~\cite{alexander1984,belloni1985,belloni1998}, enhanced counterion
localization around discrete surface charges~\cite{henle2004}, and
counterion-mediated attraction of like-charged
objects~\cite{guldbrand1984,marcelja1992,ha1997}. The latter phenomenon is
believed to be responsible for the condensation of DNA in multivalent
electrolytes~\cite{bloomfield1998}, as well as the adhesion of cement
plaste~\cite{guldbrand1984,jonsson2005}. A key part of the physics is the
attraction between an ion and its \textquotedblleft correlation
hole\textquotedblright\ resulting from a fluctuation, which has recently been
incorporated into a modified PB equation for a flat wall~\cite{santangelo2006}%
. In all of these problems, however, the equilibrium surface charge is
typically small (up to a monolayer of ions); it would be interesting to study
electrostatic correlations at a much more highly charged surface, such as an
electrode applying a large voltage, $V\gg\psi_{T}$ (our focus here).

Finally, we mention solvent effects, which are much less studied, and surely
also very important at large voltages. Electrochemists widely believe that
water dipoles in the Stern layer are so highly aligned by large electric
fields that the effective permittivity drops by an order of magnitude (e.g.
from 80$\varepsilon_{0}$ to 5$\varepsilon_{0}$)~\cite{bockris_book}. At large
applied voltages, where typical fields are of order V/nm, it is reasonable to
expect that the reduced permittivity would extend into the diffuse layer. This
could have a major effect on ion-ion correlations, since the Bjerrum length
$l_{b}\propto\varepsilon^{-1}$ could get as large as 10 nm. Other aspects of
water structure, such as hydrogen bonded networks, could also be altered by
large electric fields and large ion concentrations. It would be interesting to
perform \textit{ab initio} quantum-mechanical simulations of highly charged
double layers to investigate such effects, beyond the primitive model.

\subsection{ Scope of the present work}

In spite of the considerable literature on MPB descriptions of electrolytes in
\textit{equilibrium} or in \textit{steady state} conduction, we are not aware
of any attempt to go beyond dilute solution theory (PNP equations) in
analyzing the \textit{dynamics} of electrolytes in response to time-dependent
perturbations, such as AC voltages. Accordingly, here we develop only some
very simple models with the goal of identifying generic new features of
diffuse-charge dynamics in a concentrated solution. As such, it is preferrable
to start with equations that capture the essential physics, while remaining
analytically tractable, at least in some representative cases. For this
reason, we focus on mean-field theories of steric effects and specifically
build on the MPB equation of Borukhov et al., which can be integrated
analytically in a few simple
geometries~\cite{borukhov1997,borukhov2000,borukhov2004}. Such models also
make reasonable predictions across the entire range of voltages.

The contribution is broken into two parts. Here, in Part I, we consider the
canonical problem of charging a thin double layer, viewed as an effective
circuit element~\cite{bazant2004}. We begin in section~\ref{sec:models} by
describing two simple models for steric effects in a quasi-equilibrium diffuse
layer: (i) A ``composite layer'' model, consisting of a dilute PB outer region
and, at high voltage, an inner ``condensed layer'' of ions packed at the
steric limit, and (ii) the MPB model of Borukhov et al., which decribes the
same transition in a continuous fashion. In section~\ref{sec:circuit}, we then
analyze the diffuse-layer capacitance and its role in equivalent circuit
approximations. In section~\ref{sec:beyond}, we calculate steric effects on
salt adsorption and tangential surface conduction by the diffuse layer, and
discuss how these high-voltage effects affect and modify the applicability of
circuit models. In section~\ref{sec:compact}, we also briefly discuss the
effects of a compact dielectric layer (which could model a Stern layer of
adsorbed ions, or, more accurately, a thin film coating the solid), before
concluding in section~\ref{sec:conc}.

In Part II, we consider explicitly time-dependent problems with a general
formalism, not only applicable to thin double layers. We start with the free
energy functional of Borukhov et al. and derive modified Poisson-Nernst-Planck
equations (mPNP), based on a simple generalization of chemical potential
(\ref{eq:mu_dilute}) for concentrated solutions. As an illustration, we then
repeat the nonlinear asymptotic analysis of Ref.~\cite{bazant2004} for the
response of a blocking electrochemical cell (no reactions at the electrodes)
to a suddenly applied voltage, to expose some general consequences of steric
effects at large voltage. We also clarify the range of validity of the
thin-double-layer circuit approximations derived here in part I.


\section{Two models of steric effects in a thin diffuse
layer\label{sec:models}}


We focus on the response of a thin diffuse layer to an applied voltage, where
it suffices to consider only quasi-equilibrium properties~\cite{bazant2004},
as we justify carefully in Part II. Following Gouy and Chapman, we consider
the case of a symmetric $z:z$ electrolyte here, although our reasoning is
readily extendible to the general case. We also assume that the permittivity
$\varepsilon$ is constant in space, which is certainly not correct when dense
layers of ions form close to the surface. However, this can be taken into
account in a following step and does not change the qualitative picture
emerging from the following analysis. We will return to this point below in
section~\ref{sec:compact}.

There are at least three important lengths in our models. The first is the
Debye length $\lambda_{D}$ given by (\ref{debye}), which sets the width of the
diffuse layer at low voltage and low bulk concentration, $c_{0}$. The second
is the mean spacing of ions in the bulk electroylte, $l_{0} = (2c_{0})^{-1/3}%
$, and the third is the mean spacing of ions (essentially of the same sign) at
the maximum concentration, $a = c_{max}^{-1/3}$. A fourth scale $L$ would
characterize the geometry, as in Part II, but here we consider the regime of
thin double layers, where $\lambda_{D} \ll L$. A fifth scale would be the
Bjerrum length $l_{B}$, which we neglect by making the usual mean-field approximation.

From the first three lengths, there are two dimensionless groups. The first is
$a/\lambda_{D}$, which we assume to be small for simple electrolytes in
somewhat dilute solutions, so that steric effects are important only very
close to the surface, at the inner portion of the diffuse layer (and even
then, only at large voltages). The second dimensionless group can be written
as the mean volume fraction of ions in the bulk,
\begin{equation}
\nu= 2 a^{3} c_{0} = (a/l_{0})^{3},
\end{equation}
a natural measure of non-diluteness, which controls the importance of steric
effects, along with the dimensionless voltage, $ze\Psi_{D}/kT$. In the figures
below, we display results for $\nu=0.00005, 0.005$, and $0.5$ to span the
range from dilute to highly concentrated solutions.

\begin{figure}
[ptb]
\begin{center}
\includegraphics[
height=2.6152in,
width=3.4411in
]%
{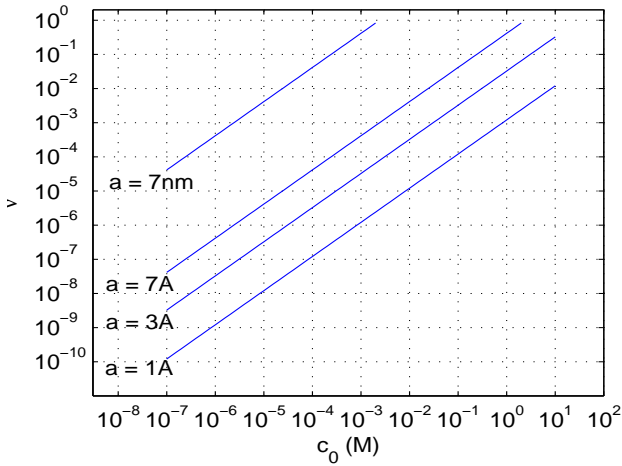}%
\caption{The dependence of the dimensionless parameter $\nu$ as a function of
the bulk concentration $c_{0}$ and effective ion size $a.$ }%
\label{fig:a_c0}%
\end{center}
\end{figure}

We stress that the phenomenological parameter $a$ is not necessarily the
diameter of an ion, $a_{0}\approx1\mathring{A}$. We prefer to think of it as a
cutoff for the unphysical divergences of PB theory, which we know must exist,
and our goal is to understand its consequences. This cutoff length could
include at least a solvation shell, e.g. $a\approx3\mathring{A}$, but ion-ion
correlations could effectively increase it further, since the Bjerrum length
is at least $7\mathring{A}$. As noted above, taking into account the decrease
of permittivity (by as much as a factor of ten) or other solvent effects at
large fields could make $l_{b}$, and thus perhaps also $a$, as large as 10 nm.
As a guide to using our results, we refer to Fig.~\ref{fig:a_c0} for the value
of the dimensionless parameter $\nu$ for different values of $a$ and $c_{0}$.%

\subsection{A composite diffuse-layer model}

\begin{figure}
[ptb]
\begin{center}
(a)\\
 \includegraphics[
height=1.4131in, width=3.4411in ]{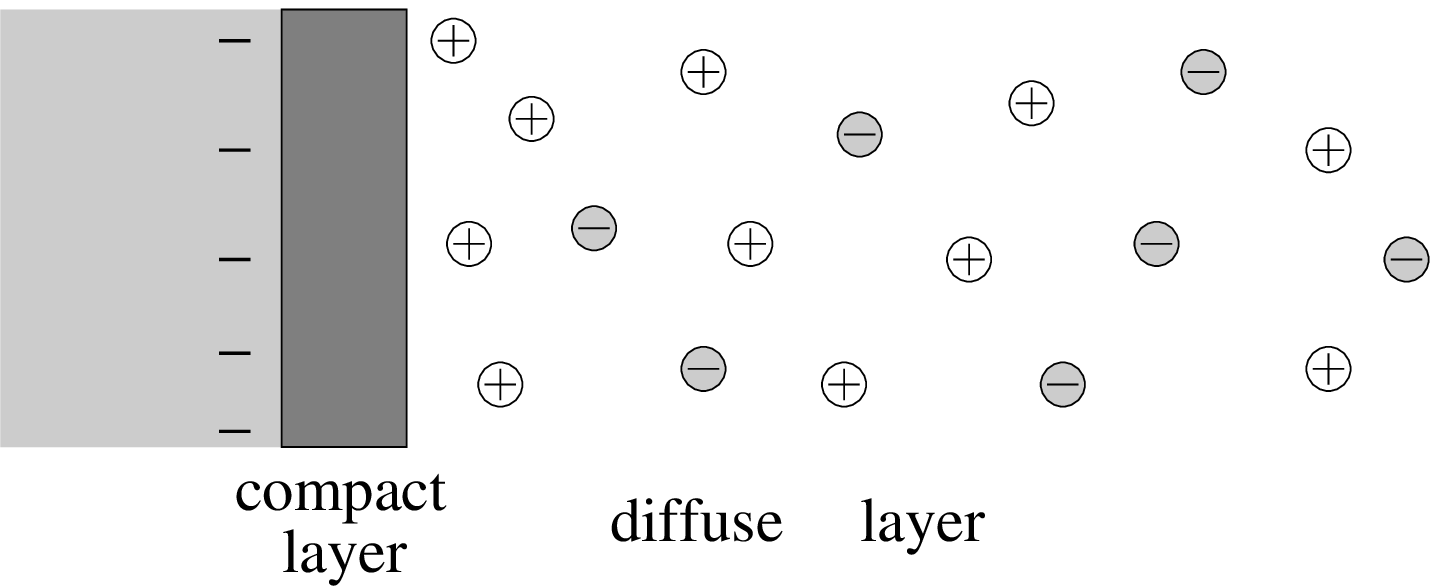}
\\
(b) \\
\includegraphics[
height=1.4131in, width=3.4411in
]%
{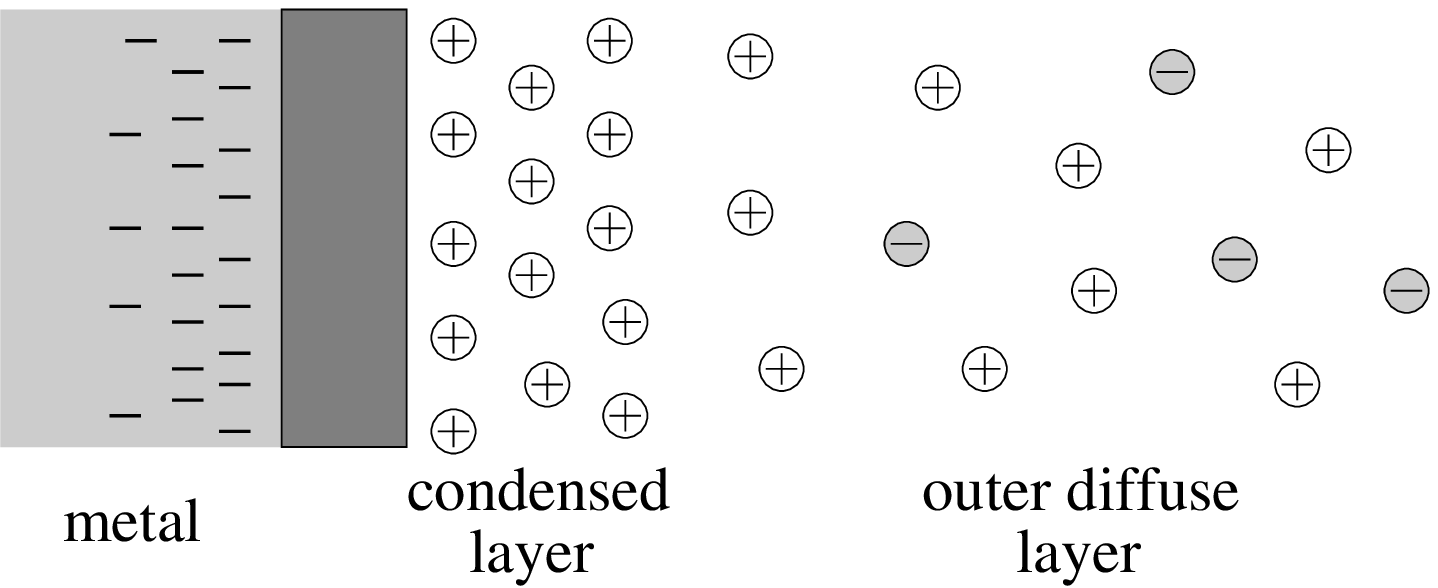}%
\caption{(a) Most prior work in electrokinetics has dealt with
  surfaces of pre-existing equilibrium charge at the scale of one
  electron per surface atom or less. This charge can be screened by
  roughly a monolayer of ions (partly in the diffuse layer) which
  corresponds to a ``small'' double layer voltage, of order $\psi_T =
  kT/e$.  (b) In contrast, nonlinear electrokinetics deals with
  polarizable (mainly metal) surfaces, where much higher surface
  charge densities can be produced by an applied voltage or electric
  field nearby, and thus the double layer can ``overcharge'' to the
  point where diluite-solution theory no longer applies. The existence
  of a minimum ion spacing implies the formation of a condensed layer
  of counterions near the surface. }%
\label{fig:pbcdl}%
\end{center}
\end{figure}

In this model, we assume that the concentration fields are governed by
Boltzmann distributions
\begin{equation}
c_{\pm}=c_{0}e^{\mp ze\psi/kT} \label{B}%
\end{equation}
wherever they are meaningful, that is, whenever these concentrations are
weaker than a physical limit $1/a^{3},$ which is set by the ion size. For both
ion species, if the formula (\ref{B}) yields a quantity bigger than $1/a^{3},$
we set the concentration field of the counter-ion to be $1/a^{3}$, and assume
that the coions are excluded from the corresponding condensed layer. The basic
physics is shown in Fig.~\ref{fig:pbcdl}.

For most geometries, such as rough surface, this model is ill-posed and would
require additional constraints to determine the location the sharp free
boundary separating the \textquotedblleft dilute\textquotedblright\ Boltzmann
region and the \textquotedblleft condensed\textquotedblright\ region at the
maximum density. For a flat, semi-infinite diffuse layer with a given total
charge $q_{D}$ or total voltage $\Psi_{D}$, however, the transition occurs at
a single well defined position, $x=l_{c}$, away from the solid surface (at
$x=0$). In that case, we separate the diffuse-layer into two parts if the
potential is strong enough
\begin{equation}%
\begin{array}
[c]{ll}%
c_{\pm}=c_{0}e^{\mp ze\psi/kT} & \mbox{ if }a^{3}c_{\pm}<1,\text{ }x>l_{c}\\
c_{\pm}=a^{-3},c_{\mp}=0 & \mbox{ if }a^{3}c_{0}e^{\mp ze\psi/kT}%
\geq1,0<x<l_{c}%
\end{array}
\label{mconcentrations}%
\end{equation}
Note that in this simplistic dichotomy the concentration field for co-ions is
slightly discontinuous at the critical potential mentioned earlier,
\begin{equation}
\Psi_{c}=\frac{kT}{ze}\ln\frac{2}{\nu}%
\end{equation}
where again $\nu=2a^{3}c_{0}$ is volume fraction occupied by (all) ions at
zero potential in the bulk. Although very simple, this model already captures
steric effects at large voltages to a great extent.%

Let us compute the thickness $l_{c}$ of the corresponding layer of "condensed"
counter-ions. We consider an electrode to which a strong negative potential
$\Psi_{D}$ is applied to the diffuse layer with respect to the bulk, such that
$|\Psi_{D}|=-\Psi_{D}>\Psi_{c}$ which leads to a condensed layer of positive
ions in its vicinity.

Poisson's equation for the thin diffuse layer reads
\begin{equation}
\varepsilon\frac{d^{2}\psi}{dx^{2}}=-\rho\label{pois}%
\end{equation}
since (for now) the permittivity $\varepsilon$ is assumed constant. Within the
condensed layer, we have $\rho=zec_{+}=ze/a^{3}$, so by integrating, we
obtain
\begin{equation}
\frac{d\psi}{dx}=-\frac{ze}{\varepsilon a^{3}}x+\frac{q_{cl}}{\varepsilon
},\text{ }\psi=-\frac{1}{2}\frac{ze}{\varepsilon a^{3}}x^{2}+\frac{q_{cl}%
}{\varepsilon}x
\end{equation}
where $-q_{cl}$ is the surface charge density on the electrode, so that
$q_{cl}$ is the total charge per unit area in the diffuse layer.

\begin{figure}
[ptb]
\begin{center}
\includegraphics[
height=2.9265in,
width=3.4411in
]%
{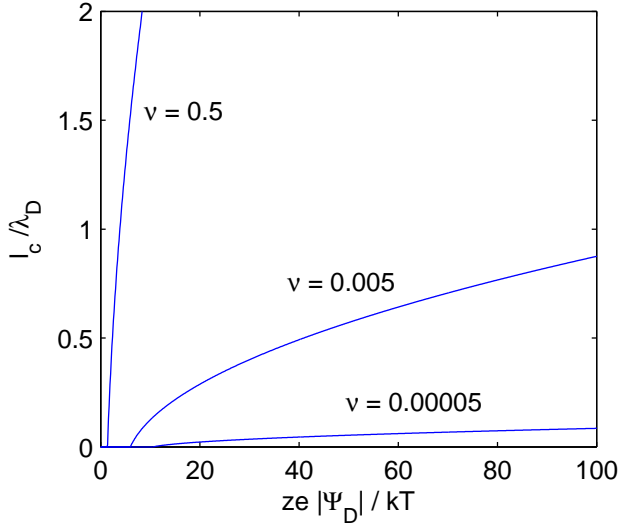}%
\caption{The thickness of the condensed layer thickness $l_{c}$ as a function
of the total voltage drop $\Psi_{D}$ across the diffuse and the condensed
layers. }%
\label{fig:Lc}%
\end{center}
\end{figure}

Within the outer diffuse layer, the standard PB equation holds,
\begin{equation}
\varepsilon\frac{d^{2}\psi}{dx^{2}}=2zec_{0}\sinh\left(  \frac{ze\psi}%
{kT}\right)
\end{equation}
which gives
\begin{equation}
\frac{d\psi}{dx}=-2\sqrt{\frac{2kTc_{0}}{\varepsilon}}\sinh\left(
\frac{ze\psi}{2kT}\right)
\end{equation}
At the interface between the condensed layer and the diffuse layer, we require
the continuity of the electric field,
\begin{equation}
q_{cl}=2zec_{0}\lambda_{D}\left(  \sqrt{\frac{2}{\nu}}-\sqrt{\frac{\nu}{2}%
}\right)  +\frac{ze}{a^{3}}l_{c}, \label{qs}%
\end{equation}
and of the electrostatic potential, so that%
\begin{equation}
-\Psi_{c}=-\frac{kT}{ze}\ln(2/\nu)=-|\Psi_{D}|-\frac{1}{2}\frac{ze}%
{\varepsilon a^{3}}l_{c}^{2}+\frac{q_{cl}}{\varepsilon}l_{c}%
\end{equation}
Combining these equations and solving for $l_{c}$ yields
\begin{equation}
l_{c}=\lambda_{D}\sqrt{2\nu}\left\{  -1+\frac{\nu}{2}+\sqrt{(1-\frac{\nu}%
{2})^{2}+[\frac{ze|\Psi_{D}|}{kT}-\ln(2/\nu)]}\text{ }\right\}  \label{LC}%
\end{equation}
which is plotted in Fig.~\ref{fig:Lc} for several values of $\nu$. Generally,
the condensed layer forms when the diffuse layer voltage $\Psi_{D}$ becomes
only a few times the thermal voltage $kT/ze$, and then it grows sublinearly,
proportionally to the square root of the potential drop as anticipated from
Poisson's equation with a constant charge density.%

From this equation and (\ref{qs}), we finally obtain the charge/potential
relation for the composite diffuse layer:%
\begin{equation}
q_{cl}=2zec_{0}\lambda_{D}\sqrt{\frac{2}{\nu}}\sqrt{(1-\frac{\nu}{2}%
)^{2}+[\frac{ze|\Psi_{D}|}{kT}-\ln(2/\nu)]} \label{qcondensed}%
\end{equation}
which holds for $|\Psi_{D}|>\Psi_{c}$, as assumed here. For weaker potentials,
there is no condensed layer and the standard PB model holds:
\begin{equation}
q_{pb}=-4zec_{0}\lambda_{D}\sinh\left(  \frac{ze\Psi_{D}}{2kT}\right)
\label{qpb}%
\end{equation}
We note that, when compared to the PB charge density $q_{pb}$, the
composite-layer charge density $q_{cl}$ is significantly reduced by the steric
effects at higher $\Psi_{D}.$ In particular, it increases only sublinearly, as
opposed to exponentially in the PB model.

\subsection{The modified PB model}

The second model we consider is the classical mean-field description of steric
effects in equilibrium mentioned in the introduction; we refer the reader to
the literature for its derivation and statistical mechanical
assumptions~\cite{wicke1952,freise1952,eigen1954,iglic1994,kralj-iglic1996,bohinc2001,bohinc2002,borukhov1997,borukhov1997,borukhov2000,borukhov2004}%
. For a symmetric $z:z$ electrolyte, the concentrations in the diffuse layer
as a function of the electrostatic potential with respect to the bulk $\psi$
are given by the modified Boltzmann distribution
\begin{equation}
c_{\pm}=\frac{c_{0}e^{\mp ze\psi/kT}}{1+2\nu\sinh^{2}\left(  \frac{ze\psi
}{2kT}\right)  } \label{MB}%
\end{equation}
where the packing parameter $\nu=2a^{3}c_{0}$ is again the bulk ion density
scaled to its maximum value and $a$ is the effective size of the ions and the
solvent molecules. Note that the concentration of each ion saturates and
cannot exceed the steric limit.

In a mean-field approximation with these ion concentrations, the potential
satisfies the modified Poisson-Boltzmann (MPB) equation,
\begin{equation}
\nabla^{2}\psi=\frac{zec_{0}}{\varepsilon}\frac{2\sinh\left(  \frac{ze\psi
}{kT} \right)  }{1+2\nu\sinh^{2}\left(  \frac{ze\psi}{2kT}\right)  }.
\label{MPB}%
\end{equation}
Unlike the composite layer model, the MPB model can be applied to any geometry
(just like the PB model). In the case of a flat diffuse layer, it gives a
similar description, except that steric effects enter smoothly with increasing
voltage, and there is no sharply defined condensed layer.

\begin{figure}
[ptb]
\begin{center}
\includegraphics[
height=2.7121in,
width=3.4402in
]%
{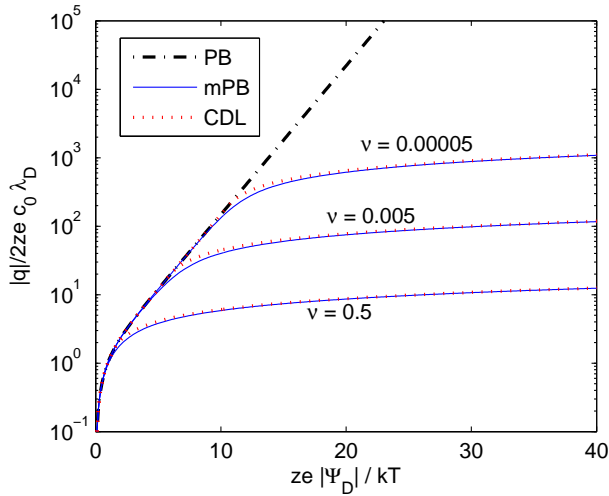}%
\caption{The diffuse layer charge given by PB, MPB and CDL models as a
function of the across potential drop.}%
\label{fig:Qd_log}%
\end{center}
\end{figure}

As for the first model, we can integrate the MPB equation across a thin double
layer to obtain the normal electric field at the inner part of the diffuse
layer
\begin{equation}
\frac{d\psi}{dx}=-\mbox{sgn}(\psi)\frac{2zec_{0}\lambda_{D}}{\varepsilon}%
\sqrt{\frac{2}{\nu}\ln[1+2\nu\sinh^{2}\left(  \frac{ze\psi}{2kT}\right)  ]}
\label{elc}%
\end{equation}
Integrating (\ref{pois}) using (\ref{elc}), we obtain for this model the
relation between the charge per unit area in the diffuse layer $q_{mpb}$ and
the potential drop across it, $\Psi_{D}$:
\begin{equation}
q_{mpb}=-\mbox{sgn}(\Psi_{D})2zec_{0}\lambda_{D}\sqrt{\frac{2}{\nu}\ln
[1+2\nu\sinh^{2}\left(  \frac{ze\Psi_{D}}{2kT}\right)  ]} \label{q}%
\end{equation}
This formula is illustrated in Fig.~\ref{fig:Qd_log}, and compared to the
analogous formula for the composite diffuse-layer model (\ref{qcondensed}).
Although the MPB concentrations (\ref{mconcentrations}) have been analyzed
carefully by previous
authors~\cite{iglic1994,kralj-iglic1996,bohinc2001,bohinc2002,borukhov1997,borukhov1997,borukhov2000,borukhov2004}%
, it seems the charge-voltage relation (\ref{q}) has been overlooked, or at
least is not stated explicitly in these papers. (Similarly, Chapman was the
first to explicitly write down the formula (\ref{qpb}) for the total
charge~\cite{chapman1913}, even though Gouy had thoroughly analyzed the
concentration and potential profiles in the \textquotedblleft Gouy-Chapman
model\textquotedblright\ a few years earlier~\cite{gouy1910}.)\newline

We note in passing that the asymptotic behaviour at large voltages is similar
for the two models presented, and corresponds to a layer of essentially
constant charge density, which results using Poisson's equation in a total
charge proportional to the square root of the potential drop.

\bigskip%

\section{Equivalent circuit elements \label{sec:circuit}}

\subsection{A word on circuit models}

As widely recognized in electrochemistry
\cite{sluyters1970,macdonald1990,geddes1997}, it is often appropriate to
describe the effect of the double layers on the dynamic properties of a system
through effective resistances and capacitances. The underlying assumption, is
that the equilibration time of the double layer is fast compared to the
dynamics of the global "circuit" considered, essentially because it is so thin
so that transport processes within the double layer are rapid. A mathematical
justification, starting from the simple PNP equations, can be given in terms
of an asymptotic analysis of the thin double layer limit~\cite{bazant2004}.

Therefore, we anticipate that for the many situations where the double layer
is thin compared to the system size, the dynamics can be understood to a
significant extent using the \emph{equilibrium characteristics} of the double
layer. These then provide appropriate boundary conditions for the dynamic
transport processes in the bulk.

In particular, the analysis presented in Ref. \cite{bazant2004}, shows that
the capacitance, or more precisely the differential capacitance
\textquotedblleft$C$\textquotedblright, of the double layer is a central
quantity allowing the modeling of the system in terms of an \textquotedblleft
RC\textquotedblright\ circuit. The second quantity of relevance, as it
characterizes the entrance into the strongly non-linear regime, is the neutral
salt uptake by the double layer which can result in an appreciable depletion
of the bulk, leading to modifications of its conductivity (affecting the
resistance \textquotedblleft$R$\textquotedblright\ in the circuit) and thus of
the dynamics. This paper and a subsequent one~\cite{chu2006} also pointed out
(in the context of dilute solution theory) that the tangential conduction
through the diffuse layer is intimately tied to neutral salt adsorption, and
indeed is governed by the same dimensionless \textquotedblleft Dukhin
number\textquotedblright.

Therefore, we will proceed to compute all of these dynamical quantities for a
thin quasi-equilibrium double layer, focusing on the general consequences of
steric effects, which are common to the two models. After that, we will return
to the question of surface capacitance and consider the effect of a thin
dielectric layer (such as an oxide coating on a metal electrode, or perhaps a
frozen Stern layer of adsorbed ions) on the overall dynamical response of the
double layer.

\subsection{Total and differential capacitances}

\begin{figure}
[ptb]
\begin{center}
\includegraphics[
height=2.6861in,
width=3.4411in
]%
{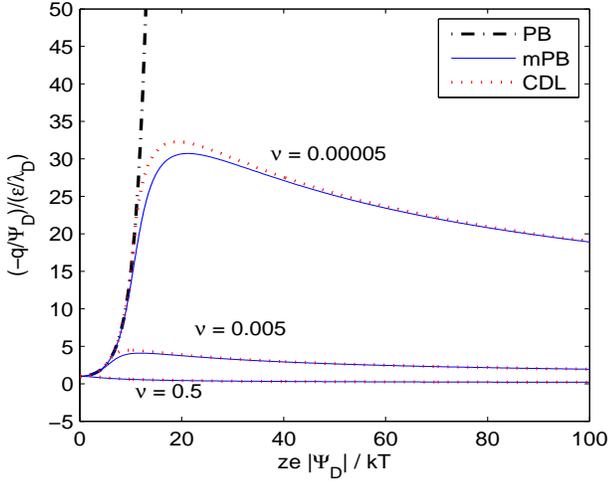}%
\caption{The total capacitance $-q(\Psi_{D})/\Psi_{D}$ of the diffuse layer as
a function of $\Psi_{D}.$}%
\label{fig:Cqd}%
\end{center}
\end{figure}

The total capacitance of the diffuse layer can be obtained directly from the
previous equations relating $q$ (the charge per unit area in the diffuse
layer) to $\Psi_{D}$ (the voltage drop across the double layer), as simply
$-q(\Psi_{D})/\Psi_{D}$. We have already computed these quantities above, and
they are compared to the PB result ($\nu=0$) in Fig.~\ref{fig:Cqd}. It is
immediately obvious that the capacitance is greatly reduced at high voltage,
compared to the predictions of dilute solution theory. The effect is so
dramatic that the capacitance in both models reaches a maximum not much larger
than the zero-voltage value, and even \textit{decreases } for all values of
voltage at sufficiently high concentration -- the opposite trend from PB theory.%

\begin{figure}
[ptb]
\begin{center}
\includegraphics[
height=2.8063in,
width=3.4411in
]%
{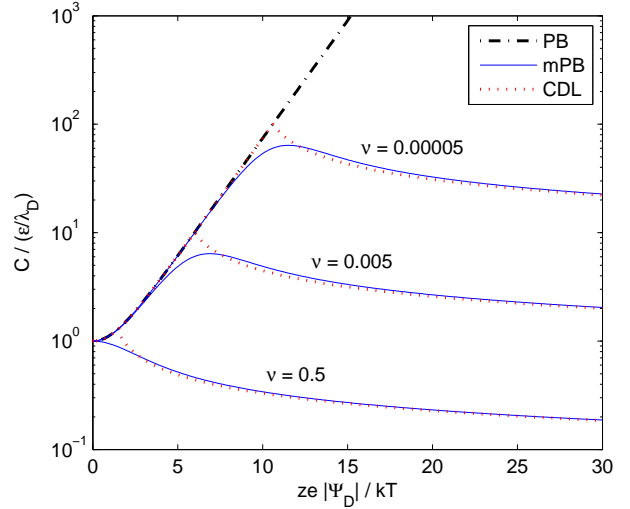}%
\caption{The diffuse layer differential capacitance as a function of the
potential drop across itself. }%
\label{fig:Cd_log}%
\end{center}
\end{figure}

\begin{figure}
[ptb]
\begin{center}
\includegraphics[
height=2.6991in,
width=3.4411in
]%
{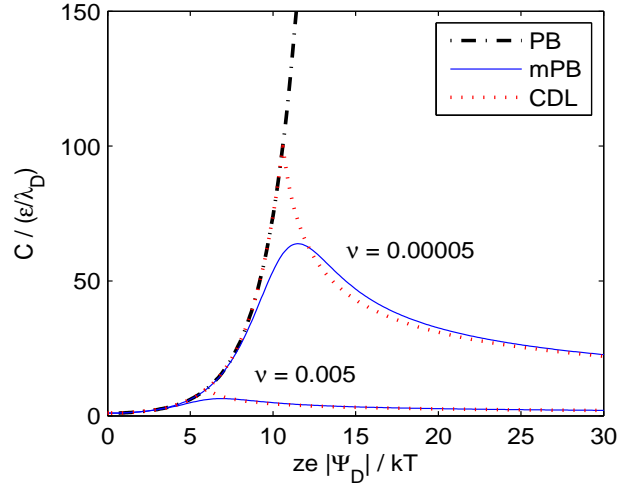}%
\caption{Same as fig.\ref{fig:Cd_log}, except with a linear scale on the
y-axis. The case $\nu=0.5$ is not shown. }%
\label{fig:Cd}%
\end{center}
\end{figure}

As noted above, the differential capacitance, defined for the diffuse layer
as
\begin{equation}
C_{D}(\Psi_{D})=-\frac{dq}{d\Psi_{D}} \label{cd}%
\end{equation}
is the relevant quantity for the dynamical response to an applied voltage.
Throughout this paper, to be clear, we will use the notation $C$ only for the
differential capacitance. In the PB model ($\nu=0$), the differential
capacitance from Eq.~(\ref{qpb}) has already been noted above:
\begin{equation}
C_{D}^{\nu}=\frac{\varepsilon}{\lambda_{D}}\cosh\left(  \frac{ze\Psi_{D}}%
{2kT}\right)  . \label{cdpb}%
\end{equation}
For the composite diffuse layer (CDL) model introduced above, (\ref{qs})
yields
\begin{equation}
C_{D}^{\nu}=\frac{\varepsilon}{\lambda_{D}}\frac{1}{\sqrt{2\nu}\sqrt
{(1-\frac{\nu}{2})^{2}+[\frac{ze|\Psi_{D}|}{kT}-\ln(2/\nu)]}} \label{CLD}%
\end{equation}
when $\Psi_{D}>\Psi_{c}.$ Otherwise, the PB formula (\ref{cdpb}) still holds.
For the MPB diffuse layer model, using (\ref{q}), we find
\begin{equation}
C_{D}^{\nu}=\frac{\frac{\varepsilon}{\lambda_{D}}|\sinh(\frac{ze\Psi_{D}}%
{kT})|}{[1+2\nu\sinh^{2}\left(  \frac{ze\Psi_{D}}{2kT}\right)  ]\sqrt{\frac
{2}{\nu}\ln[1+2\nu\sinh^{2}\left(  \frac{ze\Psi_{D}}{2kT}\right)  ]}}
\label{cdnu}%
\end{equation}
The different models are compared in Fig.~\ref{fig:Cd_log} and
Fig.\ref{fig:Cd} .%

%

The qualitative trends in $C$ are again similar in both models of steric
effects and dramatically different from PB theory. For both models, the
differential capacitance increases at very low potentials (comparable to the
thermal voltage, $kT/ze$) as it does in the PB model because steric effects
are still negligible. These enter the picture at larger potentials and limit
the storing capacity of the layer, with a differential capacitance that
actually decreases to zero at large potentials. As a consequence both models
predict a non-monotonic differential capacitance, and show a maximum at
intermediate values.

Of course, there are some clear differences in the detailed predictions of the
two models, shown in Figs.\ref{fig:Cd_log} and \ref{fig:Cd}. Although the
limiting behaviors at large and small voltage are similar, the transition is
unphysically sudden in the CDL model, compared to the more reasonable, smooth
transition in the MPB model. This is especially true at low concentrations,
where the sudden, delayed appearance of steric effects in the CDL model gives
rise to a sharp cusp in the differential capacitance versus voltage. At high
concentrations, where the \textquotedblleft low voltage\textquotedblright%
\ regime of dilute solution theory effectively vanishes, the CDL model also
fails to predict the immediate onset of steric effects, even at small
voltages, which causes a monotonic decrease in differential capacitance -- the
opposite trend of the PB model. Nevertheless, the CDL model manages to
approximate the trends of the MPB model well, with an appealingly simple
physical picture.

In summary, three basic features show up in both models, which we take as an
indication that they qualitatively hold irrespective of the specific
approximations embedded in each model: (i) the differential capacitance
$C_{D}(\Psi_{D})$ is weaker at moderate and high potentials than if steric
effects are neglected (as in the PB scheme); (ii) at moderate concentrations,
the differential capacitance varies non-monotonously with a peak at
intermediate voltages and a slow decrease towards zero at higher voltages,
(iii) at the steric limit ($\nu=1$), the differential capacitance is a
strictly decreasing function of voltage (in the MPB model). These effects are
all explained by the strong tendency of ions to form a condensed inner layer
at high voltage and/or high concentration, when steric effects are taken into
account. This greatly reduces the differential (and total) capacitance
compared to classical PB theory, which neglects the finite size of ions, and
thus predicts an absurd exponential pile-up of ions extremely close to the
surface (less than one molecular radius) in the nonlinear regime.

\subsection{ Diffuse-layer charging dynamics}

\begin{figure}
\begin{center}
(a)\includegraphics[height=3.0969in,width=3.4411in]{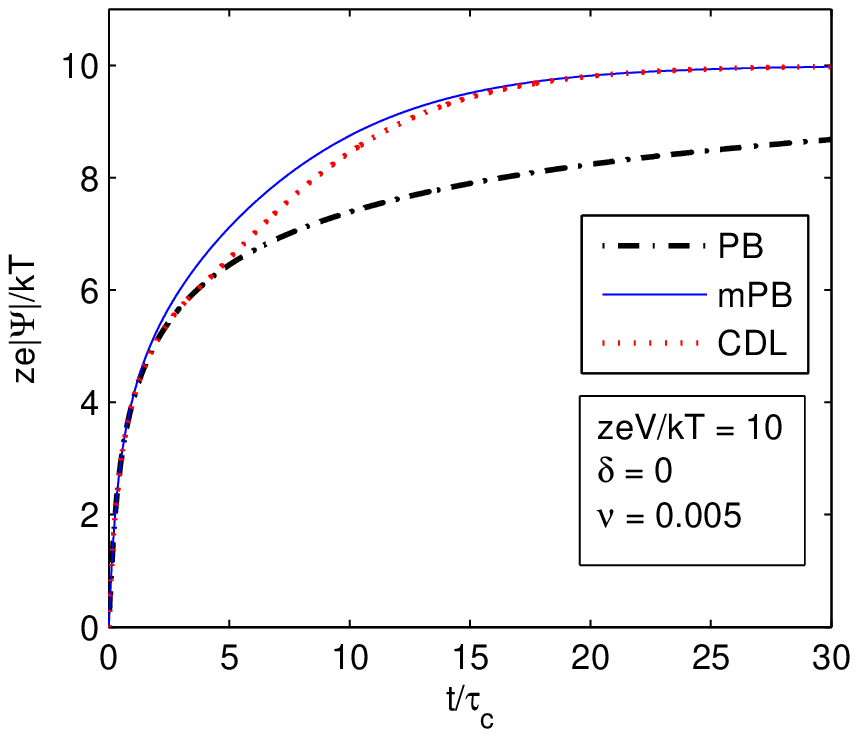} \\
(b)\includegraphics[height=3.1237in,width=3.4411in]{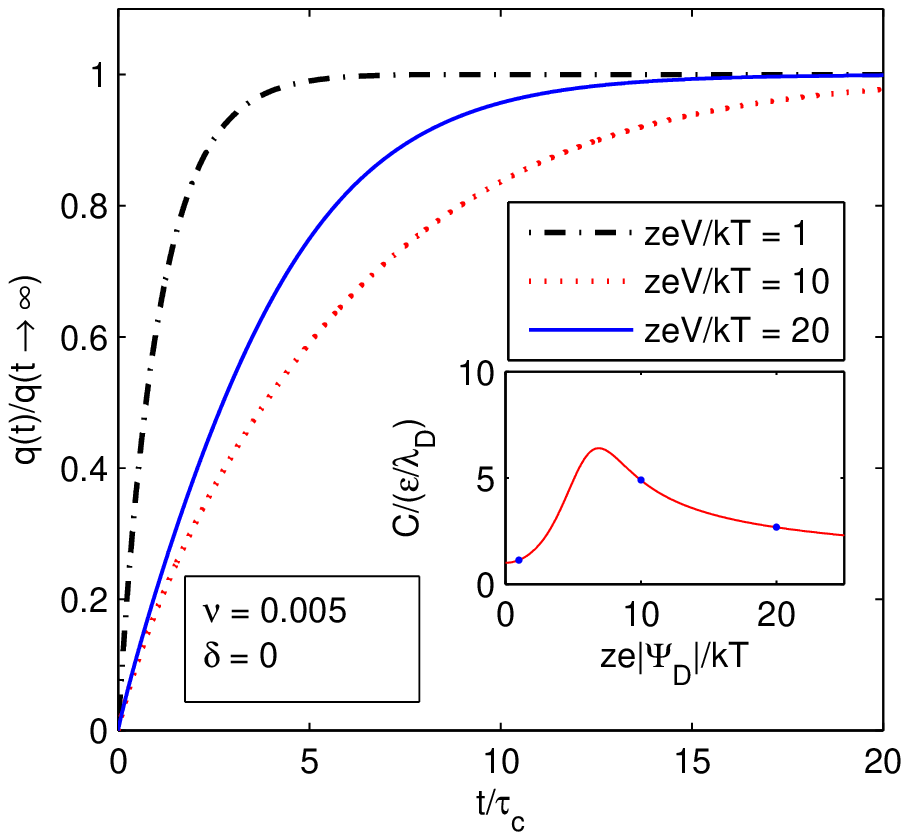}
\caption{Diffuse-layer relaxation in response to a sudden voltage
step across a blocking cell of width $2L$ and constant resistance
$R$. Time is scaled with the charging time
$\tau_{c}=\lambda_{D}L/D=RC(\Psi_D=0)$ ($\delta=0$). (a)
Relaxation of the diffuse-layer voltage $\Psi_D$ in the PB, MPB,
and CDL models for the same applied voltage, $V = 10kT/ze$. (b)
Relaxation of the total diffuse-layer charge for different
voltages, $zeV/kT = 1, 10, 20$, in the MPB model with $\nu=0.005$,
showing varying response times due to the non-monotonic voltage
dependence of the differential capacitance (inset).  }%
\label{fig:relaxnonmono}%
\end{center}
\end{figure}

These revelations have important consequences for our understanding of
double-layer charging in many situations. In the simplest picture of an
equivalent RC circuit, with $R$ the resistance of the bulk, these statements
relate to the response time of the system to a step or an AC applied voltage.
The typical response time for a driving of amplitude $V$ is $\tau
_{c}(V)=RC_{D}(V)$, so the classical picture from PB theory (\ref{cdpb}) has
been that nonlinearity greatly slows down the charging
dynamics~\cite{macdonald1954_b,simonov1977,bazant2004,chu2006,iceo2004b,olesen2006}%
. Although this may occur in a dilute solution for a relatively small range of
voltages (typically only several times the thermal voltage), steric effects in
a concentrated solution bound the relaxation time at a value much less than
expected from the PB model, in both the CDL and MPB models. This is clear in
Fig.~\ref{fig:relaxnonmono}, where we solve the RC circuit dynamics
\begin{equation}
C_{D}(\Psi_{D})\frac{d\Psi_{D}}{dt}=\frac{V-\Psi_{D}}{R} \label{eq:dyn}%
\end{equation}
for the three models numerically, to obtain the diffuse-layer voltage
$\Psi_{D}(t)$ in response to a suddenly applied voltage $V$ across the layer
in series with a bulk resistance $R$. In addition, importantly, the response
time of an electrolytic cell is found to decrease with the amplitude of the
applied voltage above the threshold for strong steric effects, $V>\Psi_{c}$.
As shown in Fig.\ref{fig:relaxnonmono}(b), the relaxation is faster for
$V=20kT/ze$ than for $10kT/ze,$ since $\Psi_{c}=8.3kT/ze$ for $\nu=0.005.$

\bigskip

\section{ Beyond circuit models \label{sec:beyond}}


\subsection{Diffuse layer salt adsorption}

As recently pointed out in Ref.~\cite{bazant2004}, circuit models can break
down at large voltages if a large number of ions (mostly counter-ions) are
engulfed by the diffuse layers with a resulting depletion of ions in the bulk.
The total salt concentration in the diffuse layer (counter-ions plus co-ions)
increases with voltage, regardless of the sign of the charge. Therefore, a
diffusion layer forms and relaxes into the neutral bulk whenever a voltage is
applied across a double layer at a blocking surface (although reactions and/or
rejection of adsorbed ions from the Stern layer could lead to negative
adsorption or salt expulsion in other situations~\cite{lyklema2005}). In the
absence of salt injection by the surface, the \textit{positive} adsorption of
neutral salt by the diffuse layer is present in the PB description, where the
counter-ion concentrations increase exponentially with
voltage~\cite{bazant2004}. It is still present but obviously weaker in models
accounting for steric effects, which severely limit the capaticity of the
diffuse layer to store additional ions at high voltage. We now quantify this
statement using the two simple models introduced above.

Following notations introduced in Ref.~\cite{bazant2004}, we define the excess
neutral salt in the double layer as \qquad%
\[
w=w^{\nu}=\int_{surface}^{bulk}(c_{+}+c_{-}-2c_{0})dx.
\]
For the PB model one finds~\cite{bazant2004},
\begin{equation}
w^{\nu=0}=8c_{0}\lambda_{D}\sinh^{2}\left(  \frac{ze\Psi_{D}}{4kT}\right)  .
\label{w0}%
\end{equation}
For the CDL model, the same equation holds for $\Psi_{D}<\Psi_{c}$, while
above this value $\Psi_{D}>\Psi_{c}$ we obtain
\begin{equation}
w^{\nu}=\left(  \frac{2}{\nu}-2\right)  c_{0}l_{c}+2c_{0}\lambda_{D}\left(
\sqrt{\frac{2}{\nu}}+\sqrt{\frac{\nu}{2}}-2\right)  \label{wcl}%
\end{equation}
with $l_{c}(\Psi_{D})$ to be extracted from (\ref{LC}). For the MPB model
\begin{equation}
w^{\nu}=\int_{0}^{\frac{ze\Psi_{D}}{kT}}\!\!\!\!\!\!\!\text{ }\frac{(\cosh
u-1)}{1+2\nu\sinh^{2}u}\frac{2c_{0}\lambda_{D}\left(  1-\nu\right)  du}%
{\sqrt{\frac{2}{\nu}\ln\left(  1+2\nu\sinh^{2}u\right)  }} \label{wnu}%
\end{equation}
which we can compute numerically.

\begin{figure}
[ptb]
\begin{center}
\includegraphics[
height=2.8184in,
width=3.4411in
]%
{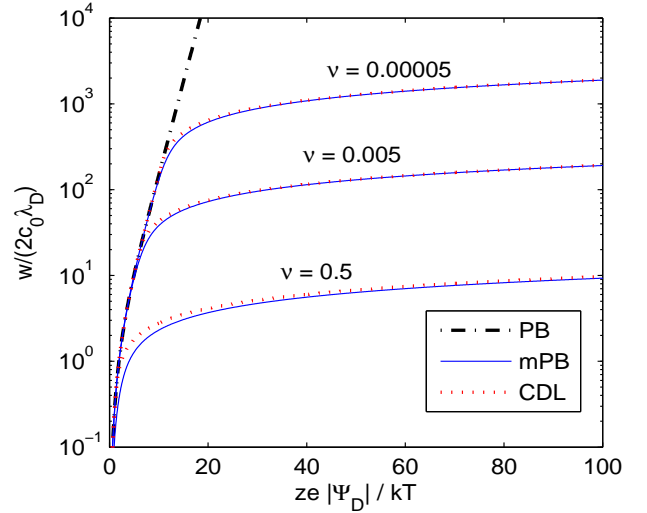}%
\caption{The diffuse layer neutral salt uptake $w$ as a function of the
potential difference $\Psi_{D}.$}%
\label{fig:Wd_log}%
\end{center}
\end{figure}

Unlike the PB formula (\ref{w0}), which predicts exponentially diverging salt
uptake for increasing $\Psi_{D},$ the steric modified formulae (\ref{wcl}) and
(\ref{wnu}) predict sublinear (square-root like) dependence on $\Psi_{D},$ at
large voltages, as can be seen by inspection of (\ref{w0})-(\ref{wnu}) or from
the plots in Fig.~\ref{fig:Wd_log}. This can be understood qualitatively as a
consequence of the roughly square-root voltage dependence of the condensed
layer width (due to its constant charge density), since most of the adsorbed
ions are condensed counterions at large voltages. (See Fig.~\ref{fig:Lc} and
Eq.~(\ref{LC}) for the CDL model.) Another way to see it in both models is
that the total salt adsorption is asymptotic to the total charge, $w\sim
q/ze$, which is true for any nonlinear double layer model (PB, MPB, CDL,...)
since the coion density goes to zero across most of the diffuse layer at large
voltages. As explained in Ref.~\cite{bazant2004}, in response to an applied
voltage the neutral bulk electrolyte becomes depleted, as the electric field
draws counterions into the diffuse layer and conducts coions away through the
bulk, resulting in a slowly expanding diffusion layer, reducing the bulk
concentration accordingly.\bigskip%

The main feature, again common to both models, is that the salt uptake at
large voltage is greatly reduced in comparison to the exponential growth
predicted by the PB picture. This is important for the dynamics as this
quantity sets the limit of applicability of the widely used RC circuit model
\cite{bazant2004}. This equivalent circuit approximation should thus hold up
to much larger values of potential when steric effects are included.

\subsection{Breakdown of circuit models}

With analytical expressions for total salt adsorption by the diffuse layer, we
can estimate the upper limits on the applied voltage where the circuit
approximation breaks down in the various models. For an electrolyte cell of
thickness $2L$, the salt uptake by the diffuse layer corresponds to a removal
of $2w$ charge carriers (ions) per unit surface, from a bulk that contained
$2c_{0}L$ such carriers initially. As long as the ratio of these two
quantities $w/(c_{0}L)$ is small, the total resistance of the RC circuit is
roughly unaffected by the salt adsorption. So, an estimate for the limit
is~\cite{bazant2004}
\begin{equation}
\alpha_{s}=\frac{w}{c_{0}L}\ll1, \label{alphas}%
\end{equation}
which translates an upper bound on the applied voltage, $|V|<V_{threshold}$
for the RC description to remain valid (in the thin double layer limit
$\lambda_{D}/L\ll1$). For the PB model, the upper bound,
\[
V_{threshold}\approx\frac{2k_{B}T}{ze}\ln\left(  \frac{L}{4\lambda_{D}%
}\right)  ,\ \ \mbox{(dilute)}
\]
is not much larger than the thermal voltage, due to the weak logarithmic
dependence on $L/\lambda_{D}$.

For the models accounting for steric effects, however, the upper bound is
greatly increased in concentrated solutions,
\[
V_{threshold}\approx\frac{k_{B}T}{ze}\left(  \frac{L}{\lambda_{D}}\right)
^{2}a^{3}c_{0}\ \ \mbox{(non-dilute)}
\]

This shows that the widely used circuit approximation for a thin double layer
$\lambda_{D}\ll L$ does not break down until enormous voltages, $V\propto
(L/\lambda_{D})^{2}$, in a concentrated solution. In a very dilute solution,
where $\nu=2a^{3}c_{0}\ll1$, the circuit approximation may break down at
moderate voltages, but only in a microsystem, where the double layers are not
so thin. For example, even in an (aqueous) electrolyte with $\lambda_{D}=10$
nm, $c_{0}=10^{-3}$ M, $a=5\mathring{A}$ and in a microdevice with $L=100\mu$m
features, the threshold voltage (with steric effects) is roughly $0.2$ Volts.

This estimate, however, neglects the possibility of a transient breakdown of
the RC circuit approximation, prior to diffusive relaxation across the entire
cell. In the case of response to a suddenly applied DC voltage, the salt
adsorption by each the diffuse layer occurs over a time scale, $\tau
_{c}=\lambda_{D}L/D$, during which diffusion spreads the nearby region of
depleted neutral bulk solution over a distance $\sqrt{D\tau_{c}}=\sqrt
{\lambda_{D}L}$. Therefore, the requirement that the local bulk conductivity
does not change significantly during charging dynamics yields the refined
estimate~\cite{bazant2004},
\[
\alpha_{d}=\frac{w}{c_{0}\sqrt{\lambda_{D}L}}\ll1
\]
which replaces $L/\lambda_{D}$ by $\sqrt{L/\lambda}$ in the estimates above.
This sets a lower bound for the limiting voltage for the validity of circuit models.

In the case of AC forcing at frequency $\omega$, the bound $\alpha_{d} \ll1$
is appropriate for low frequencies, $2\pi\omega\tau_{c}\ll1$, but circuit
models remain valid up to higher voltages at higher frequencies. At moderate
frequencies, $2\pi\omega\tau_{c} > 1$, the double layer does not have enough
time for complete charging, and the Warburg-like diffusion layer due to salt
adsorption (which oscillates at twice the frequency) only propagates to a
distance, $\sqrt{D/4\pi\omega}$. Therefore, we may crudely estimate
$(w/c_{0})\sqrt{4\pi\omega/D} \ll1$ to avoid significant changes in bulk
conductivity in the diffusion layer. (A more careful estimate would take into
account that only partial charging and salt adsorption occur with increasing
frequency.) At higher frequencies, $\omega\lambda_{D}^{2}/D \approx1$, the
diffuse layer does not have enough time to equilibrate, and little charging occurs.

These arguments can be made more precise using matched asymptotic expansions
to describe the thin double layer limit, starting from an explicitly
time-dependent model. For the case of one-dimensional response to a suddenly
applied voltage, this was done in Ref.~\cite{bazant2004} starting from the
Poisson-Nernst-Planck equations of (time-dependent) dilute solution theory. In
part II, we will derive modified PNP equations for dynamics with steric
effects and repeat the same kind of asymptotic analysis to reach a similar
conclusion our simple arguments here: Steric effects greatly extend the range
of applicability of RC circuit models, compared to what would be expected on
the basis of dilute solution theory.

\subsection{Diffuse-layer surface conduction}

Another feature not present in circuit models is the possibility of current
passed along the surface through the diffuse layer, demonstrated by
Bikerman~\cite{bikerman1933} and considered extensively in theories of
electrokinetics by Dukhin and collaborators~\cite{dukhin1993}. As first shown
in Ref.~\cite{bazant2004} and elaborated in Ref.~\cite{chu2006} in the setting
of dilute solution theory, the relative strengths of tangential
\textquotedblleft surface fluxes\textquotedblright\ through the diffuse layer,
compared to bulk fluxes, are controlled by the same dimensionless groups that
govern ion adsorption (discussed above). This is actually quite a general
result, as we now briefly explain. We will thus conclude that steric effects
also greatly reduce the importance of surface conduction in the diffuse layer
compared to the classical predictions of dilute solution theory.

Assuming small local perturbations from thermal equilibrium, the flux density
(number/area$\cdot$time) of ionic species $i$ is given by
\begin{equation}
F_{i} = - b_{i} c_{i} \nabla\mu_{i} \label{bulkflux}%
\end{equation}
where the chemical potential $\mu_{i}$ generally has a different form than
(\ref{eq:mu_dilute}) in a concentrated solution (e.g. see Part II). Consider a
thin diffuse layer near a charged surface, where the ion concentration $c_{i}$
departs from its nearby neutral bulk value $c_{i}^{b}$. Due to fast relaxation
at the small scale of the screening length, the diffuse-layer concentration
remains in quasi-equilibrium at nearly constant chemical potential in the
normal direction, $\mu_{i} \sim\mu_{i}^{b}$, in the thin double layer limit.
There can, however, be small tangential gradients, $\nabla_{\|} \mu_{i}\neq0$
at the macroscropic length scale leading to an excess diffuse-layer
``surface'' flux density (number/length$\cdot$time):
\begin{align}
F_{i}^{s}  &  =\int_{surface}^{bulk} (-b_{i} c_{i} \nabla_{\|} \mu_{i} + b_{i}
c_{i}^{b} \nabla_{\|} \mu_{i}^{b}) dx\\
&  \sim-\nabla_{\|} \mu_{i}^{b} \, \int_{surface}^{bulk} b_{i}(c_{i}-c_{i}%
^{b})\, dx
\end{align}
For a constant mobility $b_{i}$, this takes the same form as the bulk flux
density (\ref{bulkflux}),
\begin{equation}
F_{i}^{s} = -b_{i} \Gamma_{i} \nabla_{\|} \mu_{i}%
\end{equation}
where the bulk concentration (number/volume) has been replaced by the
diffuse-layer ``surface concentration'' (number/area),
\begin{equation}
\Gamma_{i} = \int_{surface}^{bulk} (c_{i}-c_{i}^{b})\, dx
\end{equation}
In a concentrated solution, we generally expect that the mobility $b_{i}$
might decrease in the difuse layer, due to steric effects and large normal
electric fields, so this formula may overestimate the surface flux density.

Following Bikerman~\cite{bikerman1933,bikerman1940,chu2006}, we may estimate
the relative importance of surface to bulk flux densities at a length scale
$L$ by the dimensionless group
\begin{equation}
\frac{F_{i}^{s}}{F_{i}^{b} L} = \frac{\Gamma_{i}}{c_{i}^{b} L}%
\end{equation}
which we see also measures the relative importance of ``surface adsorption''
of ion $i$ in the diffuse layer relative to the bulk concentration. For a
highly charged diffuse layer $|\Psi_{D}|\gg\Psi_{c}$, the ions are mostly of
one type (counter-ions), so $\Gamma_{i} \sim q \sim w$. The surface current is
also carried mostly by those ions, $J^{s} \sim ze F_{i}^{s}$, while the bulk
current is carried by both ions, $J^{b} \approx2(ze) F_{i}^{b}$ (neglecting
diffusion compared to electromigration). Therefore, we see that the ``Dukhin
number'' comparing surface conduction at a highly charged diffuse layer to
bulk conduction at nearly uniform concentration,
\begin{equation}
\mbox{Du} = \frac{J^{s}}{J_{b} L} \sim\frac{w}{2c_{0} L} = \frac{\alpha_{s}%
}{2}%
\end{equation}
is roughly half of the dimensionless group $\alpha_{s}$ governing salt
adsorption by the diffuse layer.

We have seen that steric effects greatly reduce $w$ compared to the
predictions of dilute solution theory, so that $\alpha_{s} = O(\lambda_{D}/L)
$ remains small up to rather high voltages. Since the calculation above
over-estimates the importance of surface conduction, it is clear that steric
effects also greatly reduce the Dukhin number compared to the predictions of
dilute solution theory. We conclude that surface conduction in a thin diffuse
layer does not become important until voltages large enough to violate the
equivalent circuit approximation are applied across it. Compact layer surface
conduction may still be important in some cases, but it too is limited by the
same steric effects.

\section{Compact Layer Effects\label{sec:compact}}

We now check the robustness of these conclusions to the additional presence of
an insulating surface layer between the metal electrode and the electrolyte
where the EDL models are applied. We suppose that this layer is not involved
in the dynamics so that its properties do not change, and assume for the sake
of simplicity that these properties are linear so that it can be described as
a fixed surface capacitance $C_{s}$.

Such models have been used in many circumstances, sometimes to describe the
Stern layer corresponding to condensed ions. Our approach here is different in
the sense that a layer of condensed ions would be involved in the
charging/discharging process. What we have in mind is closer to a thin film,
e.g. of oxide on the metal, of thickness $h_{s}$ and dielectric constant
$\epsilon_{s}$ so that $C_{s}=\epsilon_{s}/h_{s}$. This form has been proposed
to model coating layers on electrodes in the context of AC
electrokinetics~\cite{ajdari2000}.

The overall differential capacitance of the interface is now
\[
\frac{1}{C}=\frac{1}{C_{D}}+\frac{1}{C_{S}}%
\]
with $C_{S}=dq/d\Psi_{S},$ where $q$ is the total charge per area in the
double layer, and $\Psi_{S}$ is the voltage drop across the aforementioned
compact layer. The total voltage drop across the interface is $\Psi=\Psi
_{S}+\Psi_{D}$, and $q/\Psi$ is the total capacitance of the interface.

A useful dimensionless parameter to quantify the effect of this surface layer
is $\delta=\varepsilon/(\lambda_{D}C_{S})$, which has been employed recently
in general studies of diffuse-charge dynamics~\cite{bonnefont2001,bazant2004},
as well as in theoretical
models~\cite{ajdari2000,gonzalez2000,ramos2003,iceo2004a,iceo2004b,olesen2006}
and in the fitting of experimental data~\cite{levitan2005,ramos2005} for
induced-charge (AC or DC) electrokinetics. No surface layer corresponds to
$\delta=0$. In the above mentioned picture of the layer of oxide,
$\delta=(\varepsilon/\epsilon_{s})(h_{s}/\lambda_{D})$ so that even very thin
oxide layers can yield not too small values of $\delta$ if $\varepsilon
/\epsilon_{s}$ is large.

\begin{figure}
[ptb]
\begin{center}
\includegraphics[
height=2.7821in,
width=3.4411in
]%
{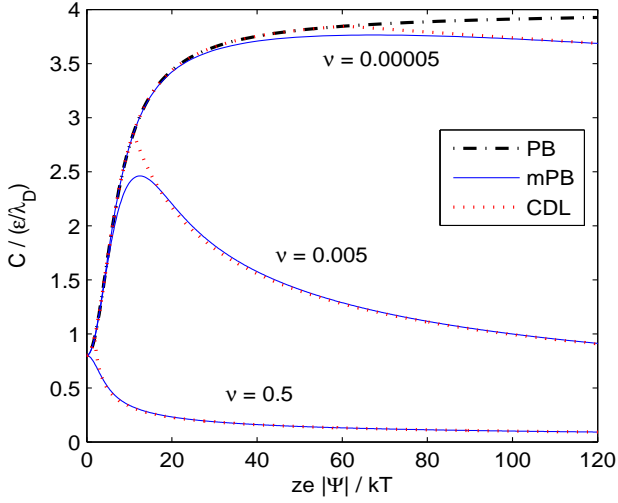}%
\caption{The total double layer differential capacitance in presence of a
Stern layer with $\delta=0.25$ versus the total potential drop across the
double layer.}%
\label{fig:Cds}%
\end{center}
\end{figure}

Let us now compare the differential capacitance $C$ (with surface layer) to
the bare double layer differential capacitance previously plotted in
Fig.\ref{fig:Cd}, for the PB equation and the two models with steric effects.
The corresponding plots are provided on Fig.\ref{fig:Cds} for $\delta=0.25$.

\bigskip%

Many qualitative points are obvious. First, the PB differential capacitance
does not blow up exponentially anymore, as the surface layer takes over when
the double layer voltage gets large, so that $C$ converges to the finite value
$C_{s}$ within the PB model. Most of the potential drop is then across the
surface layer. Second, the two main consequences of steric effects pointed
above remain valid: $C$ is weaker when steric effects are taken into account,
and $C(\Psi)$ is nonmonotonous with a maximum at intermediate values and a
further decrease to zero.%

\begin{figure}
[ptb]
\begin{center}
\includegraphics[
height=2.7051in,
width=3.4411in
]%
{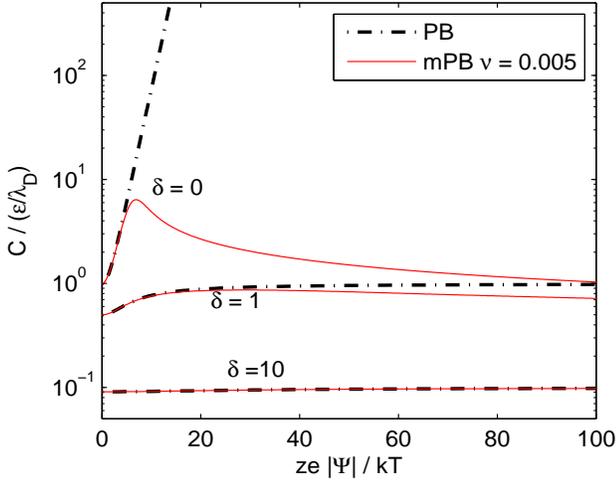}%
\caption{The differential capacitance $C$ of the double layer as a function of
potential difference $\Psi$ for various values of Stern layer capacitance
values. }%
\label{fig:Cdelta}%
\end{center}
\end{figure}
%

\begin{figure}
[ptb]
\begin{center}
\includegraphics[
height=4.2981in,
width=3.4411in
]%
{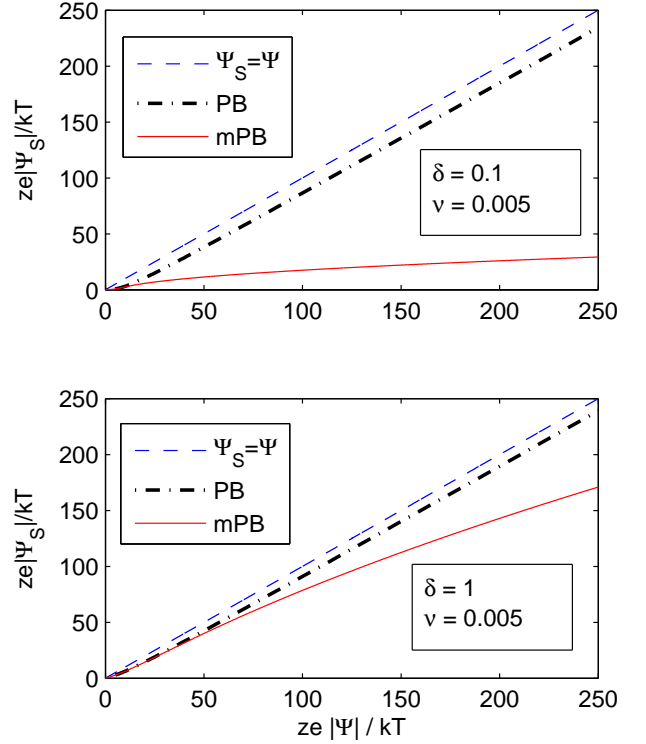}%
\caption{The Stern layer voltage drop as a function of the total double layer
voltage difference. For reference, the line $\Psi_{S}=\Psi$ is also drawn. In
the PB theory, almost all of the voltage drop is realized over the Stern layer
when the voltage drop is on the order of a few voltages. }%
\label{fig:psispsi}%
\end{center}
\end{figure}

When the parameter $\delta$ is zero or small(i.e. the Stern layer capacitance
is large), a closely packed "diffuse" layer has to form in response to a high
potential drop. This is the case when the steric effects become important and
the PB theory fails. However, for the higher values of $\delta,$ most of the
potential drop is realized over the Stern layer, and the ions in the diffuse
layer are not as densely packed. Thus the system stays below the steric limit,
and the PB theory agrees with the MPB theory, as demonstrated by the total
capacitance plots of Fig. \ref{fig:Cdelta}. Similar comments can be made on
Fig.\ref{fig:Cds}, which shows that for $\delta=0.25,$ steric effects may or
may not be important depending on value of the dimensionless parameter
$\nu=2a^{3}c_{0}.$ For the distribution of the voltage drop over the Stern
layer and the diffuse layers, see Fig.\ref{fig:psispsi}.

This analysis, although correct, is somewhat misleading, however, since the
assumption of a constant compact-layer capacitance may not always be
reasonable. In many situations of interest, where several volts ($\approx100
kT/e$) are applied across the double layer, it is unlikely that the compact
layer could withstand a significant fraction of the total voltage. Note that
dielectric breakdown in water can occur in average fields as low as 20 MV/m =
0.02 V/nm in experiments applying submicrosceond voltage pulses (well below
the charging time of the double layers)~\cite{jones1995}. The critical field
may be higher in the Stern layer, where water is confined by ions at the outer
Helmholtz plane~\cite{bockris_book}, but it seems implausible for an atomic
layer to sustain several volts without somehow ``short circuiting'' via
electron cascades, electrolysis, Faradaic reactions, etc. In some cases, the
compact layer models a dielectric thin film, which may be considerably
thicker. Again, however, most coating materials, such as teflon or various
metal oxides, undergo dielectric breakdown in fields of order 10 MV/m, so a
dielectric coating cannot easily withstand several volts unless it is at least
10 nm wide.

In general, we see that the capacitance of the compact layer must effectively
decrease at large voltages, which corresponds to the limit $\delta\to0$. As a
result, a signficant fraction of a large voltage must be sustained by the
diffuse layer, making steric effects important in many situations of interest.
Regardless of the accuracy of our simple models, therefore, we believe that
the predicted qualitative effects of ion crowding are likely to have broad
relevance for experimental systems applying large voltages.


\section{Conclusion\label{sec:conc}}


We have used two simple models for the double layer to account for crowding
effects which necessarily take place at intermediate and large applied
voltages. These models are both based on modifications of the Poisson
Boltzmann description of dilute solutions. This strategy has lead us to
identify important operational consequences of these crowding effects the thin
double layers, namely a largely reduced double layer capacitance and a
decreased ion uptake from the bulk.

We have provided in these sections explicit formulas for the total and
differential capacitance and for the salt uptake of interfaces \emph{at
equilibrium}, as a function of the potential drop across the interface. More
precisely we have recalled the PB results (no steric effects $\nu=0$), and
given results for the two simple models with steric effects ($\nu\neq0$),
considering for all cases the possibility of a finite insulating layer on the
electrode ($\delta\neq0$).

The two models lead to remarkably similar results suggesting that these
semi-quantitatively hold beyond the specifics of these models. Both showed
marked differences with the PB approach: the differential capacitance and salt
uptake are much weaker, and the former varies non-monotonously with the
applied potential.

These observations for the \emph{equilibrium} properties have led us to make
predictions for the \emph{dynamics} of electrolyte cells of size quite larger
than the Debye length: (i) an effective equivalent RC circuit description
holds for a wider range of potentials than expected on the simple basis of the
PB equation, (ii) the response time is much smaller than expected from PB at
large voltages, (iii) this time decreases at large voltages after an initial
increase for lower values.

The dramatic effect of steric constraints in this problem also shows that
other predictions of nonlinear PB theory, such as the change of scaling from
$V^{2}$ to $|V|\log|V|$ for AC electro-osmosis~\cite{olesen2006}, are limited
in applicability and should be revisited with models taking crowding effects
into account. More generally, this work suggests that, beyond the present
problem of ionic transport in electrolytic systems, the description of
electrokinetic effects at large applied voltages should be revisited to
correct shortcomings of dilute-solution theory.

\section*{Acknowledgments}

This work was supported in part by the MRSEC program of the National Science
Foundation under award number DMR 02-12383 (MZB). AA acknowledges very
gratefully the hospitality of the Department of Mathematics at MIT where this
study was realized. \bigskip

\bigskip

\bibliographystyle{plain}
\newcommand{\noopsort}[1]{} \newcommand{\printfirst}[2]{#1}
\newcommand{\singleletter}[1]{#1} \newcommand{\switchargs}[2]{#2#1}

\end{document}